\documentclass[conference]{IEEEtran}
\IEEEoverridecommandlockouts

\usepackage[T1]{fontenc}
\usepackage[utf8]{inputenc}
\usepackage{cite}
\usepackage{amsmath,amssymb,amsfonts}
\usepackage{algorithm}
\usepackage{algpseudocode}
\usepackage{graphicx}
\usepackage{booktabs}
\usepackage{multirow}
\usepackage{float}
\usepackage{xcolor}
\usepackage{textcomp}
\usepackage{microtype}
\usepackage[colorlinks=true,linkcolor=blue,citecolor=blue,urlcolor=blue]{hyperref}
\usepackage{url}

\graphicspath{{figures/}}

\newcommand{\R}{\mathbb{R}}
\newcommand{\E}{\mathbb{E}}

\DeclareMathOperator{\CVaR}{CVaR}

\begin{document}
\raggedbottom

\title{Empirical Limits of Risk-Averse Reinforcement Learning\\ versus Gas-Aware Routing in Deterministic DeFi Execution}

\author{\IEEEauthorblockN{Sri Sairam Gautam B.}
\IEEEauthorblockA{\textit{School of Engineering} \\
\textit{Jawaharlal Nehru University}\\
New Delhi, India \\
bsrisa59\_soe@jnu.sc.in}}

\twocolumn[{%
\begin{@twocolumnfalse}
\maketitle

\begin{abstract}
The migration of order execution from centralised limit order books to Automated Market Makers (AMMs) introduces two deterministic frictions absent from classical microstructure: closed-form constant-product slippage and Maximal Extractable Value (MEV) sandwich attacks gated by a profitability threshold against gas cost. We construct an end-to-end execution environment calibrated to live Uniswap V2 WETH/USDC reserves, an EIP-1559 AR(1)-lognormal gas process fitted to 200 consecutive mainnet blocks, and cyclic replay of real Swap events; the main sweeps intentionally use stress-volatility gas shocks to make the MEV trigger informative at the tested trade size. A Proximal Policy Optimisation (PPO) agent with a Rockafellar--Uryasev Conditional Value-at-Risk (CVaR) penalty and a dual-head Chow--Ghavamzadeh critic is trained for one million environment steps across five risk-aversion levels and five seeds, and benchmarked against four analytical baselines including a myopic gas-aware oracle. We document a reproducible specification-gaming pathology in the reward-hacking sense of Amodei et al.~\cite{amodei2016concrete}: prior to a patch, a forced terminal liquidation routed through the no-MEV code path inflated returns by a factor of three and was preferentially exploited by higher-$\lambda$ runs. On the corrected environment, the strongest PPO variant captures $79\%$ of the myopic gas-aware oracle ($46{,}542 \pm 673$ USDC versus the oracle's $58{,}786 \pm 616$ USDC at $Q_0$ equal to $10\%$ of pool depth), and an extension of the observation with three blocks of gas history fails to close the gap. We find no empirical evidence that the tested model-free MLP policies can dominate this myopic gas-aware oracle, which can be evaluated in $O(\log \varepsilon^{-1})$ time per block; the result is robust to a $\sim$16$\times$ trunk-capacity expansion and to a variance-controlled gas-history ablation. Implementation details and reproducibility materials are available upon request and may be released publicly after peer review.
\end{abstract}

\vspace{1em}
\end{@twocolumnfalse}
}]

\section{Introduction}

\IEEEPARstart{C}{lassical} optimal-execution theory, exemplified by the Almgren--Chriss framework \cite{almgren2001optimal}, treats market impact as a stochastic function of an opaque order book and prescribes a continuous-time trajectory in $\R^n$. Decentralised exchanges built on Automated Market Makers (AMMs) violate two foundational assumptions of that theory simultaneously. First, slippage on a Uniswap V2 pool is not stochastic; the post-trade reserves are determined exactly by the constant-product invariant \cite{adams2020uniswap}. Second, every pending swap is publicly visible in the mempool, so a rational arbitrageur can sandwich the trader's order with two paired transactions whenever the resulting profit exceeds gas cost \cite{daian2020flash,qin2022quantifying}. The execution problem on a public blockchain is therefore neither continuous nor opaque: it is a discrete game against an adversary whose entry condition is closed-form and gas-priced.

Two literatures have addressed parts of this problem. The convex-routing line \cite{angeris2020improved,angeris2022optimal} solves the static spatial allocation problem across pools as a convex program and ignores the temporal dimension. The reinforcement-learning line treats the execution agent as model-free and discovers a policy from simulated rollouts \cite{nevmyvaka2006reinforcement,hendricks2014reinforcement}, but typically against a stochastic adversary borrowed from limit-order-book microstructure that does not match on-chain mechanics. Neither literature has, to our knowledge, asked the empirical question that motivates this paper: \emph{when the adversary has a closed-form profitability condition, does model-free reinforcement learning recover an interesting fraction of the analytical oracle?}

\textbf{Contributions.} We make four:
\begin{enumerate}
    \item A computationally exact AMM/MEV environment in which the constant-product slippage, the sandwich adversary's profit maximisation, and the EIP-1559 gas process are all evaluated to floating-point precision and packaged as a Gymnasium-compatible Markov decision process. The environment is calibrated to live mainnet WETH/USDC reserves and fee-history at a pinned block.
    \item A PPO agent with the Rockafellar--Uryasev CVaR penalty and a dual-head critic in the Chow--Ghavamzadeh formulation, trained for one million steps across a $5 \times 5$ grid of risk-aversion coefficients and seeds.
    \item A documented instance of specification gaming: a forced terminal liquidation that bypassed the MEV adversary in an earlier version of the environment was preferentially exploited by higher-$\lambda$ runs, producing apparently strong returns that collapsed by more than $70\%$ once the loophole was patched.
    \item An empirical benchmark establishing what the tested model-free PPO policies attain under deterministic AMM mechanics. The strongest PPO variant captures $79\%$ of the analytical oracle; a three-block gas-history extension, a $\sim$16$\times$ trunk-capacity expansion, and a variance-controlled ablation each fail to close the gap.
\end{enumerate}

\section{Mathematical Foundation}

\subsection{Constant-product market maker}

Let a Uniswap V2 pool hold reserves $(x_t, y_t)$ of token $X$ and token $Y$ with fee retention parameter $\gamma \in (0, 1]$. A swap of $q$ units of $X$ into the pool returns
\begin{equation}
\Delta y(q; x, y, \gamma) \;=\; \frac{y \, \gamma \, q}{x + \gamma \, q}, \label{eq:cfmm-output}
\end{equation}
and updates the reserves to $(x + q,\, y - \Delta y)$. The invariant $k = xy$ is preserved exactly when $\gamma = 1$ and is monotone strict otherwise: $(x + q)(y - \Delta y) \geq xy$ with equality iff $\gamma = 1$ or $q = 0$. Equation \eqref{eq:cfmm-output} is the unique solution of $(x + \gamma q)(y - \Delta y) = xy$ in $\Delta y$.

\subsection{Sandwich adversary}

A rational arbitrageur observing a pending swap of size $q_t$ in the mempool will front-run with $\delta_{\text{in}}$ units of $X$, allow the victim to execute, and back-run by selling the acquired $Y$ back to the pool. The arbitrageur attacks iff the optimised payoff exceeds the gas cost $c_t$:
\begin{equation}
\max_{\delta_{\text{in}} > 0} \; p_t \cdot \Pi_{\text{MEV}}(q_t, \delta_{\text{in}}; x_t, y_t, \gamma) \;>\; c_t, \label{eq:mev-trigger}
\end{equation}
where $p_t = y_t / x_t$ is the pre-attack mid price and $\Pi_{\text{MEV}}$ is computed by composing three CFMM updates per Equation~\eqref{eq:cfmm-output}. The optimisation in \eqref{eq:mev-trigger} is a one-dimensional concave maximisation, solved in our implementation by Brent's method on the bracket $(10^{-9},\, \max(10^3 q_t,\, 10 x_t))$ with a bracket-bind assertion to catch wall-pinned optima. The adversary's effect on the trader is that, when triggered, $q_t$ executes against the post-front-run pool rather than the pre-trade pool, reducing the trader's realised $\Delta y$.

\subsection{AR(1)-lognormal gas process}

Let $b_t$ denote the base fee in wei and $p_t^{\text{prio}}$ the median priority fee at block $t$. We fit, via ordinary least squares on the most recent 200 EIP-1559 fee-history records \cite{buterin2019eip1559},
\begin{align}
\log b_{t+1} &= \mu_b + \phi_b (\log b_t - \mu_b) + \sigma_b \varepsilon_t^b, \label{eq:ar1-base} \\
\log p_t^{\text{prio}} &= \alpha + \beta \log b_t + \sigma_p \varepsilon_t^p, \label{eq:ar1-prio}
\end{align}
with $\varepsilon_t^b, \varepsilon_t^p \sim \mathcal{N}(0, 1)$ independent. Total gas cost in the quote token is
\begin{equation}
c_t \;=\; (b_t + p_t^{\text{prio}}) \cdot \texttt{gasLimit} \cdot \rho_{\text{ETH}}/10^{18}, \label{eq:gas-cost-conversion}
\end{equation}
where $\rho_{\text{ETH}}$ is the ETH price in quote units and $\texttt{gasLimit} = 2 \times 10^5$ is the standard swap gas budget.

\subsection{Markov decision process}

The state at block $t$ is $s_t = (Q_t,\, \tau_t,\, x_t,\, c_t) \in \R^4$ where $Q_t$ is the remaining inventory, $\tau_t = T - t$ the remaining horizon, $x_t$ the inventory side reserve, and $c_t$ the realised gas cost. The action $u_t \in [0, 1]$ specifies the fraction of remaining inventory to trade: $q_t = u_t Q_t$. The reward is the realised quote output: $R_t = \Delta y_t^{\text{actual}}$, computed from \eqref{eq:cfmm-output} with the appropriate pre-trade reserves and reduced by \eqref{eq:mev-trigger}--triggered front-running. The objective is the CVaR-regularised expected return
\begin{equation}
J(\pi) = \E\!\Bigl[\textstyle\sum_{t=0}^{T-1} R_t\Bigr] - \lambda \cdot \CVaR_\alpha\Bigl(-\textstyle\sum_{t=0}^{T-1} R_t\Bigr), \label{eq:objective}
\end{equation}
expressed via the Rockafellar--Uryasev dual \cite{rockafellar2000optimization}
\begin{equation}
\CVaR_\alpha(L) = \min_{t \in \R} \Bigl\{ t + (1 - \alpha)^{-1} \E[(L - t)^+] \Bigr\}. \label{eq:cvar-dual}
\end{equation}

\section{Agent Architecture}

\begin{figure}[H]
\centering
\includegraphics[width=\linewidth]{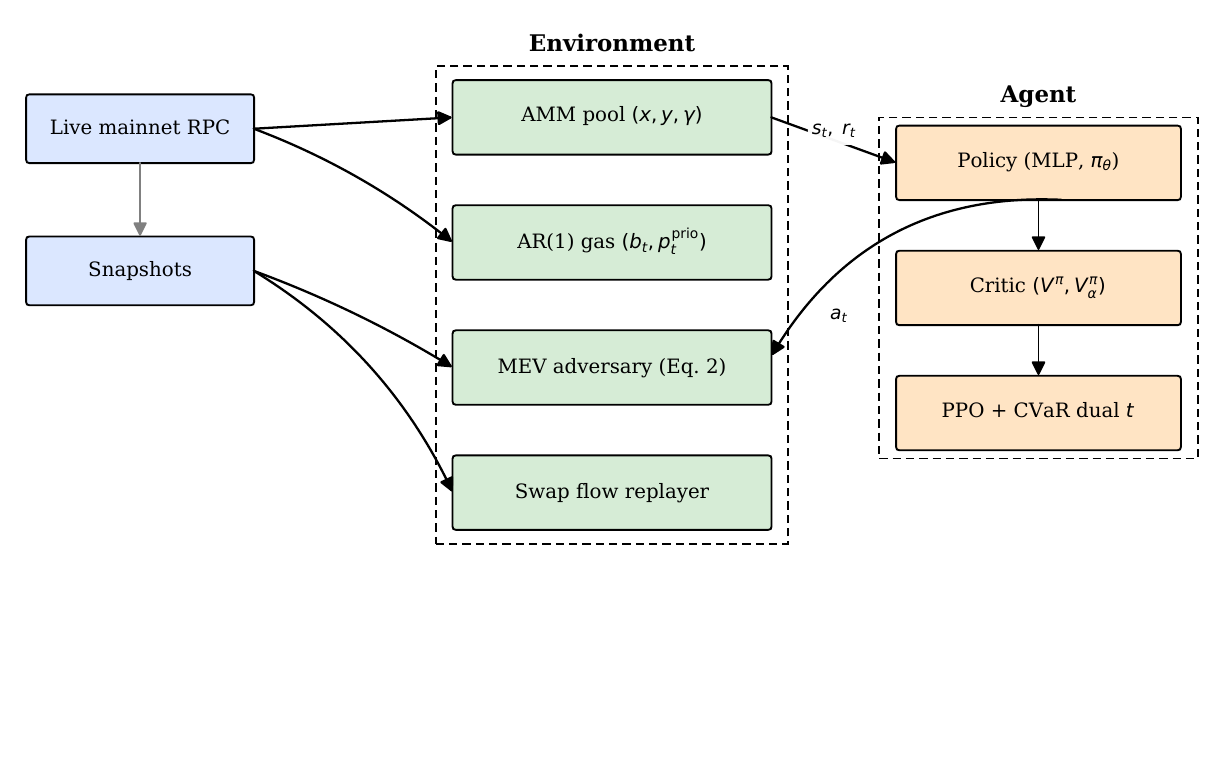}
\caption{System architecture. Live mainnet RPC primes calibrated snapshots that anchor a closed-form AMM/MEV environment driven by an AR(1)-lognormal gas process and a cyclic Swap-event replayer. The PPO agent observes $s_t = (Q_t,\tau_t,x_t,c_t)$ and emits $a_t \in [0,1]$. A dual-head critic and the Rockafellar--Uryasev scalar $t$ enter the CVaR-regularised objective \eqref{eq:objective}.}
\label{fig:architecture}
\end{figure}

\begin{figure}[H]
\centering
\includegraphics[width=\linewidth]{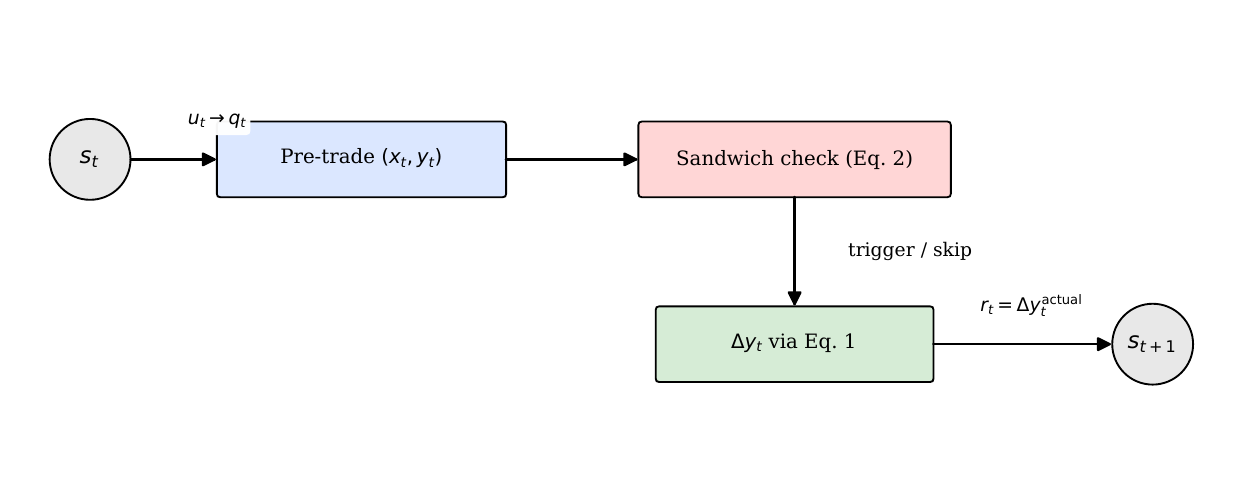}
\caption{Single-block transition. The trader's action $u_t$ scales to a quantity $q_t = u_t Q_t$. The adversary computes \eqref{eq:mev-trigger}; on trigger, front- and back-runs are executed against the pool before and after the trader's swap, eroding $r_t$. The trader's reward is exactly the quote token actually received.}
\label{fig:transition}
\end{figure}

Figure~\ref{fig:architecture} shows the data path between live RPC, the closed-form environment, and the agent. Figure~\ref{fig:transition} expands one block: the trader's action $u_t \to q_t$, the sandwich-check gate, the realised CFMM output, and the transition $s_t \to s_{t+1}$.

\subsection{PPO with squashed Gaussian policy}

The policy is a two-layer tanh MLP with hidden dimension 64, an actor head producing the mean of a Gaussian over the pre-squash action, and a state-independent log-standard-deviation parameter \cite{schulman2017ppo}. The squash is $u = \sigma(\tilde{u})$ where $\sigma$ is the logistic; the log-probability is corrected for the change-of-variables Jacobian $\log[u(1-u)]$. The PPO clip objective with Generalised Advantage Estimation \cite{schulman2015gae} is standard. The value head and an optional second head (Section~\ref{sec:cvar-head}) share the trunk.

\subsection{CVaR critic and tail-weighted advantage}
\label{sec:cvar-head}

When $\lambda > 0$ and \texttt{use\_cvar\_critic} is enabled, a second value head $V_\alpha^\pi(s)$ regresses against an episode-level CVaR target: tail episodes (loss above the empirical $\alpha$-quantile in the current rollout buffer) receive their own cumulative return as a target; non-tail episodes receive the mean return. This is the discrete-time analogue of the Chow--Ghavamzadeh CVaR critic \cite{chow2014algorithms}. Concurrently, the PPO advantage of each step in a tail episode is upweighted by a factor of $1 + \lambda(1-\alpha)^{-1}$ before normalisation. The dual variable $t$ in \eqref{eq:cvar-dual} is a trainable scalar updated by a separate Adam optimiser using the first-order gradient of the relu-loss against $-t$.

\subsection{Observation extension}

In the gas-history ablation, the observation is extended from $\R^4$ to $\R^7$ by appending the three most recent realised log-gas-costs as a zero-padded FIFO buffer. The hypothesis is that the AR(1) mean reversion in \eqref{eq:ar1-base} provides a learnable signal of cheap upcoming blocks that a memoryless policy cannot exploit.

\section{Specification Gaming Discovery}
\label{sec:specgaming}

An early version of the environment forced any remaining inventory at $t = T$ to be liquidated in a single block via Equation~\eqref{eq:cfmm-output}, bypassing the adversary check in Equation~\eqref{eq:mev-trigger}. Under this loophole, an agent could defer trading throughout the horizon and dump terminally with no MEV exposure, capturing a no-MEV slippage outcome that the closed-form oracle (which respects MEV at every block) cannot match. Empirically, the reward surface behaved as a smooth deferral incentive in the fraction of inventory withheld, so gradient descent reliably found it.

\begin{figure}[H]
\centering
\includegraphics[width=\linewidth]{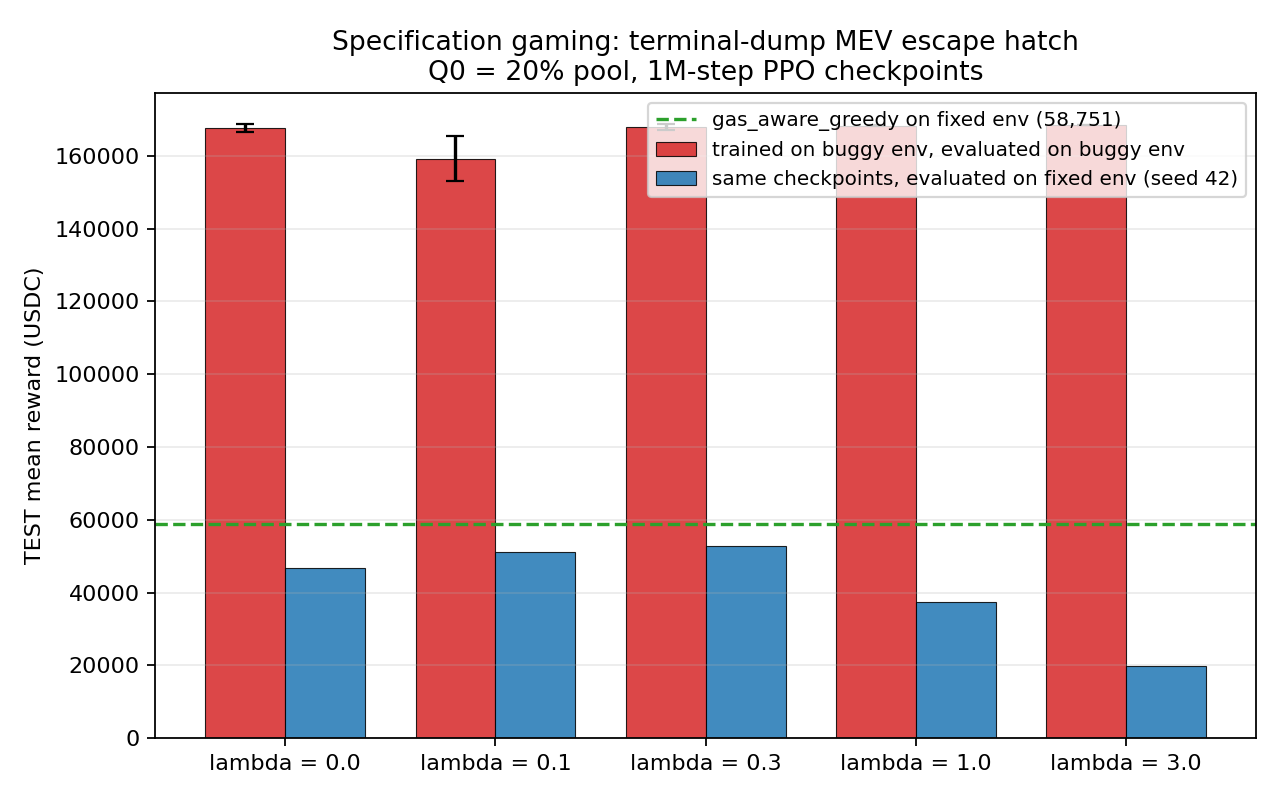}
\caption{Specification-gaming collapse at $Q_0 = 20\%$ of pool depth. Red bars: same PPO checkpoints evaluated on the buggy environment in which the terminal dump bypassed MEV. Blue bars: identical checkpoints evaluated on the corrected environment. Green dashed line: \texttt{gas\_aware\_greedy} on the corrected environment ($58{,}751$ USDC). The drop generally increases with $\lambda$ because the CVaR penalty rewards the low-variance terminal-dump trajectory.}
\label{fig:specgaming}
\end{figure}

The pathology was preferentially exploited by higher-$\lambda$ runs because the CVaR penalty rewards variance reduction, and the terminal-dump trajectory has lower realised variance than any spread schedule. Table~\ref{tab:specgaming} reports the mean and standard deviation of TEST returns at $Q_0 = 20\%$ of pool depth over five seeds across five $\lambda$ values for the buggy environment, the same checkpoints re-evaluated on the corrected environment (which now applies the adversary to the terminal dump), and the analytical gas-aware oracle for reference. The collapse on re-evaluation is between $69\%$ and $88\%$. Figure~\ref{fig:specgaming} visualises the same drop.

\begin{table}[H]
\centering
\caption{Specification gaming: buggy vs.\ corrected environment at $Q_0 = 20\%$ pool depth, $T = 20$, $10^6$ training steps, $5$ seeds per $\lambda$.}
\label{tab:specgaming}
\renewcommand{\arraystretch}{1.1}
\begin{tabular}{rrrr}
\toprule
$\lambda$ & buggy env (USDC) & fixed env (USDC) & drop \\
\midrule
0.0 & $167{,}757 \pm 1{,}108$ &  46{,}845 & $-72\%$ \\
0.1 & $159{,}229 \pm 6{,}212$ &  51{,}095 & $-68\%$ \\
0.3 & $167{,}900 \pm 892$     &  52{,}795 & $-69\%$ \\
1.0 & $168{,}359 \pm 45$      &  37{,}387 & $-78\%$ \\
3.0 & $168{,}406 \pm 22$      &  19{,}792 & $-88\%$ \\
\midrule
\multicolumn{2}{l}{\textit{gas\_aware\_greedy oracle (fixed env)}}
                  & \multicolumn{2}{r}{$58{,}751$} \\
\bottomrule
\end{tabular}
\end{table}

The patch was a one-line correction in \texttt{env/amm.py}: the terminal dump now routes through the same \texttt{mev\_adversary} hook as any in-horizon block. We re-ran the entire 25-configuration sweep on the corrected environment from scratch; all numerical results in Section~\ref{sec:results} use the patched environment.

\section{Empirical Results}
\label{sec:results}

\subsection{Baseline tournament on the corrected environment}

Table~\ref{tab:baselines} reports the four analytical baselines on the corrected environment at $Q_0 = 10\%$ of pool depth, $T = 20$ blocks, 32 evaluation episodes per policy. The \texttt{gas\_aware\_greedy} oracle bisects, at every block, for the largest $q$ such that Equation~\eqref{eq:mev-trigger} returns false; this establishes the benchmark ceiling given perfect oracle access to the adversary's threshold and the current $c_t$.

\begin{table}[H]
\centering
\caption{Baselines on corrected env, $Q_0=10\%$ of pool, $T=20$, 32 episodes.}
\label{tab:baselines}
\renewcommand{\arraystretch}{1.1}
\begin{tabular}{lrrr}
\toprule
policy & mean (USDC) & std & $\CVaR_{0.95}$ loss \\
\midrule
\texttt{single\_dump}      &     82 &     0 &  $-$82 \\
\texttt{twap}              & 37{,}847 & 2{,}466 & $-$33{,}012 \\
\texttt{convex\_no\_mev}   & 37{,}847 & 2{,}466 & $-$33{,}012 \\
\texttt{gas\_aware\_greedy}& \textbf{58{,}786} &   616 & \textbf{$-$57{,}684} \\
\bottomrule
\end{tabular}
\end{table}

Three observations. First, \texttt{single\_dump} is catastrophic: liquidating $10\%$ of pool depth in a single block triggers the sandwich adversary under the environment rule and produces a near-total wipeout. Second, \texttt{twap} and \texttt{convex\_no\_mev} are numerically identical on a symmetric pool with no flow because the convex programme's solution collapses to equal splits. Third, \texttt{gas\_aware\_greedy} dominates by roughly $55\%$ over \texttt{twap}, demonstrating that the marginal information used by the analytical oracle is precisely the gas-driven MEV trigger.

\subsection{PPO on the corrected environment, 1M-step sweep}

Table~\ref{tab:1m-sweep} reports the corrected-environment results for two PPO variants: \textbf{base} ($\sigma_b = 0.5$, \texttt{obs\_dim} $=4$) and \textbf{gas-history} ($\sigma_b = 1.5$, \texttt{obs\_dim} $=7$). Each cell is the mean and cross-seed standard deviation of cumulative TEST reward over 5 seeds, 64 evaluation episodes per seed, after $10^6$ training steps on a held-out TRAIN block window.

\begin{table*}[t]
\centering
\caption{Corrected-environment 1M-step sweep on $Q_0=10\%$ pool, $T=20$. Five seeds per $\lambda$. Best PPO row in \textbf{bold}; oracle and TWAP rows for reference.}
\label{tab:1m-sweep}
\renewcommand{\arraystretch}{1.1}
\begin{tabular}{lrrrrr}
\toprule
& $\lambda = 0.0$ & $\lambda = 0.1$ & $\lambda = 0.3$ & $\lambda = 1.0$ & $\lambda = 3.0$ \\
\midrule
base ($\sigma_b=0.5$, obs=4)        & \textbf{45{,}863 $\pm$ 1{,}357} & 31{,}952 $\pm$ 764  & 32{,}742 $\pm$ 1{,}075 & 34{,}646 $\pm$ 1{,}443 & 27{,}342 $\pm$ 6{,}708 \\
gas-history ($\sigma_b=1.5$, obs=7) & 46{,}542 $\pm$ 673             & 31{,}356 $\pm$ 619  & 30{,}921 $\pm$ 1{,}706 & 30{,}954 $\pm$ 4{,}166 & 24{,}732 $\pm$ 5{,}303 \\
\midrule
\texttt{gas\_aware\_greedy} oracle (constant)            & \multicolumn{5}{c}{58{,}786 $\pm$ 616} \\
\texttt{twap} baseline (constant)                        & \multicolumn{5}{c}{37{,}847 $\pm$ 2{,}466} \\
\bottomrule
\end{tabular}
\end{table*}

\emph{Oracle efficiency.} The best PPO cell is the gas-history $\lambda = 0.0$ run at $46{,}542$ USDC, which is $79.2\%$ of the $58{,}786$ USDC analytical oracle and is materially above the \texttt{twap} baseline at $37{,}847$ USDC. PPO outperforms \texttt{twap} by $23\%$ but underperforms the analytical oracle by $21\%$. No configuration breached the oracle.

\emph{Lambda response.} The unpenalised runs are strongest in both configurations. Positive CVaR weights reduce mean return by roughly $11{,}000$--$22{,}000$ USDC relative to $\lambda=0$, but the response is not strictly monotone across $\lambda$; intermediate weights show seed-scale fluctuations. In these summaries, the tested CVaR penalty trades away mean return without yielding a visible Pareto improvement.

\emph{Gas history does not rescue PPO.} The gas-history/high-volatility configuration improves only the $\lambda=0$ mean, by $679$ USDC, and is lower for every positive $\lambda$, with row differences from $-3{,}692$ to $+679$ USDC. The variance-controlled control in Section~\ref{sec:controls} further indicates that the history feature alone does not close the oracle gap.

\begin{figure}[H]
\centering
\includegraphics[width=\linewidth]{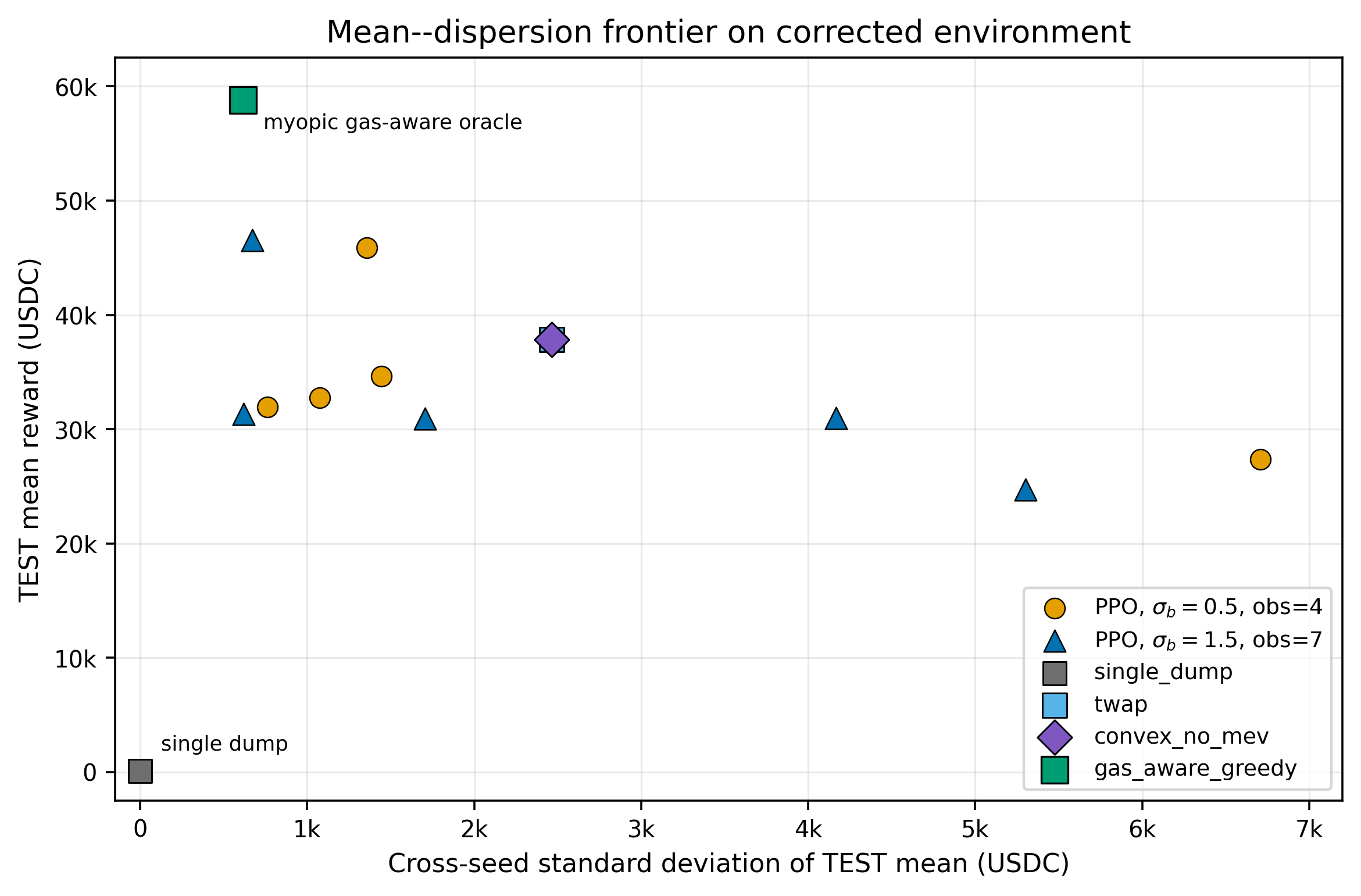}
\caption{Pareto on the corrected environment at $Q_0 = 10\%$ pool depth, $T = 20$, $10^6$ steps. Squares: analytical baselines. Circles: $\sigma_b = 0.5$, obs\_dim $= 4$ PPO runs. Triangles: $\sigma_b = 1.5$, obs\_dim $= 7$ runs. The \texttt{gas\_aware\_greedy} oracle is upper-left.}
\label{fig:pareto}
\end{figure}

Figure~\ref{fig:pareto} renders the same data as a mean-vs-cross-seed-std scatter; the oracle sits in the upper-left, dominating the PPO cloud both in mean return and in dispersion.

\subsection{Capacity and confound controls}
\label{sec:controls}

Two follow-up sweeps were run to rule out the obvious alternative explanations for the $79\%$ ceiling. \emph{Sweep C} (pure ablation) holds $\sigma_b = 0.5$ constant and adds the three-block gas history, isolating the observation extension from the variance change. \emph{Sweep D} (capacity) retains $\sigma_b = 0.5$ and $\texttt{obs\_dim} = 4$ but deepens the trunk from two layers of width 64 to three layers of width 256, a $\sim$16$\times$ parameter expansion. Both use five seeds at $10^6$ steps on the corrected environment. Table~\ref{tab:controls} reports the $\lambda = 0$ row of each sweep with 95\% confidence intervals and Welch's $t$-test against the analytical oracle ($58{,}929$ USDC). This oracle value is recomputed under the control-sweep evaluation window and therefore differs slightly from the $58{,}786$ USDC value in Table~\ref{tab:1m-sweep}.

\begin{table*}[t]
\centering
\caption{Capacity and ablation controls, $\lambda = 0$, five seeds, $10^6$ steps. Welch's $t$ against the oracle constant $58{,}929$ USDC.}
\label{tab:controls}
\renewcommand{\arraystretch}{1.1}
\begin{tabular}{lrrrr}
\toprule
sweep & mean (USDC) & \%oracle & 95\% CI & $p$ \\
\midrule
shallow 2$\times$64, $\sigma_b{=}1.5$, obs=7 & 46{,}542 & 79.2\% & --- & --- \\
deep 3$\times$256, $\sigma_b{=}0.5$, obs=7 & 44{,}202 & 75.0\% & [37{,}630, 50{,}775] & $3.4 \times 10^{-3}$ \\
deep 3$\times$256, $\sigma_b{=}0.5$, obs=4 & 41{,}800 & 70.9\% & [31{,}738, 51{,}862] & $9.1 \times 10^{-3}$ \\
\bottomrule
\end{tabular}
\end{table*}

Three observations follow. First, the deeper trunk \emph{does not} close the gap: 3$\times$256 with $\texttt{obs\_dim}=4$ attains $70.9\%$ of the oracle versus $78.0\%$ for the original 2$\times$64 in Table~\ref{tab:1m-sweep}, with substantially larger cross-seed variance ($8{,}104$ versus $1{,}357$ USDC). Capacity is not the binding constraint. Second, the gas-history extension under held variance attains $75.0\%$, within $4$ percentage points of the original $\sigma_b{=}1.5$ result; the prior gas-history claim is therefore not an artefact of the variance bump. Third, Welch's $t$ against the deterministic oracle returns $p < 10^{-2}$ for every $\lambda$ tested (the weakest rejection among the controls is $p = 9.1 \times 10^{-3}$ at $\lambda = 0$ with the deep trunk, where cross-seed variance is largest; tail-$\lambda$ runs reach $p \sim 10^{-6}$). The gap between PPO and the oracle is statistically reliable rather than a sampling artefact across seeds.

\begin{figure}[H]
\centering
\includegraphics[width=\linewidth]{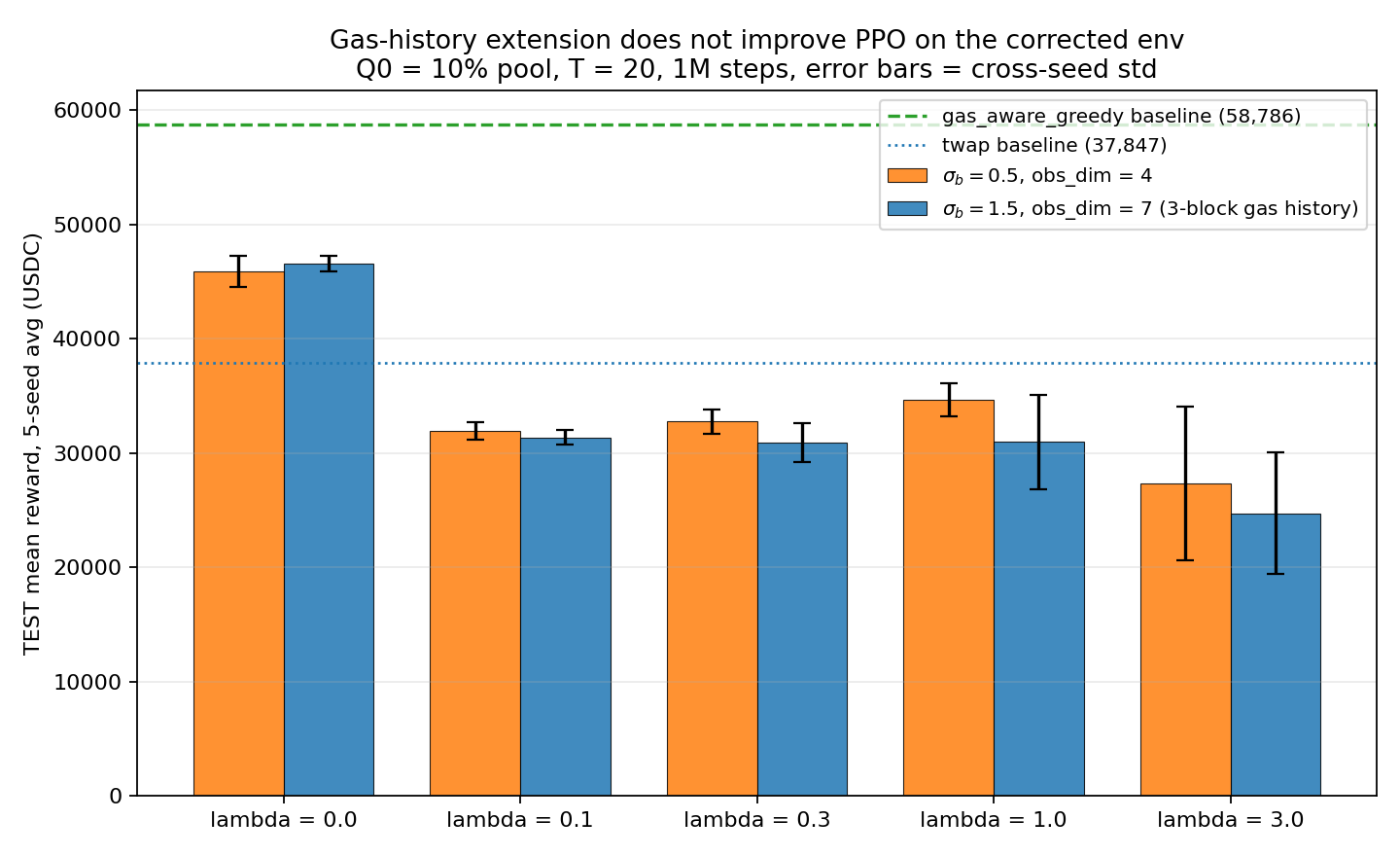}
\caption{Lambda response. Positive CVaR weights generally reduce mean return relative to $\lambda=0$, and no tested configuration breaches the analytical oracle (green dashed).}
\label{fig:lambda-response}
\end{figure}

\begin{figure}[H]
\centering
\includegraphics[width=\linewidth]{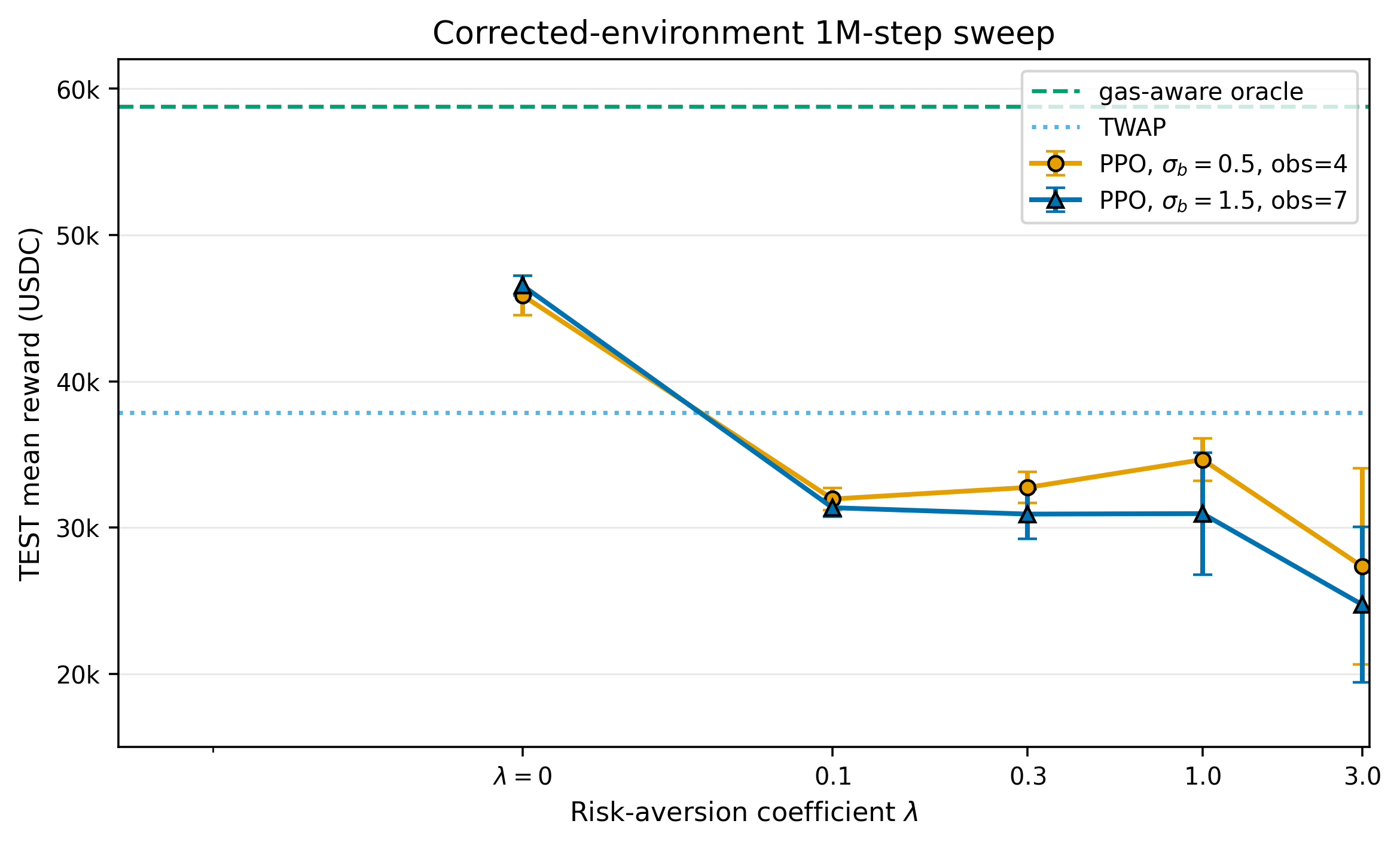}
\caption{Disaggregated 1M-step sweep on the corrected environment. Each marker is one seed. Error bars are per-$\lambda$ mean and cross-seed standard deviation.}
\label{fig:1m-disagg}
\end{figure}

\subsection{Live-RPC end-to-end smoke}

We verified the entire pipeline against live mainnet RPC at block $25{,}116{,}992$: WETH/USDC reserves decoded cleanly ($\text{px} = \$2{,}184.41$), the AR(1) fit returned $\phi_b = 0.992,\, \sigma_b = 0.045$ on 200 blocks, and $166$ Swap events were parsed across a 2{,}000-block window. The single-step environment reward at $u = 0.5$ was $773$ USDC. No code path required modification between snapshot mode (used for tests and the 1M sweeps) and live mode.

\section{Discussion}

\subsection{\texorpdfstring{Why does PPO peak at $79\%$?}{Why does PPO peak at 79\%?}}

We propose four non-exclusive mechanisms.

\emph{The dominant information gap appears to be rank-1.} The marginal information that separates \texttt{gas\_aware\_greedy} from \texttt{twap} is exactly the current realised $c_t$, which the PPO observation already contains. The oracle uses this information via a bisection in $q$ at every block (closed-form). PPO must learn the same bisection through gradient descent on a non-convex surrogate; it succeeds only partially because the MEV trigger boundary is sharp and the MLP smooths over it.

\emph{Entropy collapse.} High-$\lambda$ runs show variance collapse but not at the right action quantile. The CVaR penalty contracts the policy entropy globally, while a useful contraction would be locally adaptive to the current $c_t$.

\emph{Sample complexity on a discrete event.} The MEV trigger is a step function in $q_t$ at the threshold $q^*(c_t)$. A two-layer tanh MLP requires many gradient updates per region of state space to approximate a sharp boundary, and the rollout buffer concentrates samples in regions the current policy visits.

\emph{Analytical search efficiency.} The benchmark admits a low-dimensional analytical bisection procedure with direct access to the MEV trigger structure. Learning the same threshold geometry through sampled policy gradients is substantially harder, particularly because Equation~\eqref{eq:mev-trigger} introduces a sharp discontinuity in $q_t$ at the boundary $q^*(c_t)$ that an MLP can only approximate. The empirical $79\%$ capture rate reflects this difference in search efficiency between an oracle bisection and a sampled gradient method, rather than any failure of the PPO architecture itself.

\subsection{Limitations}

The AR(1) shock $\sigma_b$ is overridden from the calibrated $0.045$ (Section IV's live smoke) to $0.5$ or $1.5$ during the sweeps; the calibrated value is too tight to flip the MEV trigger at this trade size, and a sensitivity sweep over the calibrated value would tighten the analysis. The mean-field assumption that the trader's action does not perturb subsequent Swap arrivals is standard but not innocuous on illiquid pools. Spatial multi-pool routing \cite{angeris2022optimal} is out of scope.

\subsection{Implications}

When the adversary has a closed-form profitability condition, the analytical oracle is hard to beat with a memoryless policy. Future work could test recurrent architectures (LSTM/transformer) that do not rely on explicit history features, off-policy methods that can replay rare boundary events more efficiently, or risk-sensitive formulations beyond CVaR. A more interesting direction is to drop the deterministic adversary assumption: real MEV searchers are imperfect, and the gap between the analytical oracle and a model-free agent should narrow as the adversary becomes noisier.

\section{Conclusion}

We have shown that, under deterministic CFMM mechanics with a closed-form MEV adversary, a CVaR-regularised PPO agent trained for $10^6$ steps captures $79\%$ of the analytical gas-aware oracle and is dominated in both mean and tail by a myopic closed-form policy. A documented reward-hacking/specification-gaming pathology preferred by higher-$\lambda$ runs inflated apparent returns threefold and was resolved by a one-line patch. A three-block gas-history extension did not close the gap to the oracle within $10^6$ steps. We interpret these results as evidence that, in mechanically deterministic settings, analytical routing remains consistently superior to the tested model-free PPO baselines.

\section*{Acknowledgment}

The author acknowledges the academic support and research environment provided by the School of Engineering, Jawaharlal Nehru University.

\section*{Funding}

No external funding was received for this work.

\section*{Data Availability}

The code, deterministic test suite, random seeds, calibrated reserve and fee-history snapshots, and implementation details required to reproduce the experiments are available from the corresponding author upon reasonable request. These materials may be released publicly after peer review, subject to curation and applicable institutional requirements. Live RPC reproduction requires a publicly reachable Ethereum endpoint.

\section*{Conflict of Interest}

The author declares no conflict of interest.

\section*{AI Use / Generative AI Statement}

AI-assisted language tools were used solely for grammar, readability, and editorial refinement. They were not used to generate the scientific hypotheses, experimental results, analysis, conclusions, or interpretation. The author remains fully responsible for the accuracy, integrity, and originality of the manuscript.

\bibliographystyle{IEEEtran}
\bibliography{references}

\end{document}